\documentclass[aip,amsmath,amssymb,reprint, jcp]{revtex4-1}

\usepackage{graphicx}% Include figure files
\usepackage{dcolumn}% Align table columns on decimal point
\usepackage{bm}% bold math
\usepackage{multirow}					
\usepackage{array}
\usepackage{float}
\usepackage{mathtools}
\usepackage{color}
\usepackage{natmove}
\usepackage{lipsum}
\usepackage{listings}
\usepackage{xcolor}
\usepackage{subfig}

\lstdefinestyle{mystyle}{
    backgroundcolor=\color{backcolour},   
    commentstyle=\color{codegreen},
    keywordstyle=\color{magenta},
    numberstyle=\tiny\color{codegray},
    stringstyle=\color{codepurple},
    basicstyle=\ttfamily\footnotesize,
    breakatwhitespace=false,         
    breaklines=true,                 
    captionpos=b,                    
    keepspaces=true,                 
    numbers=left,                    
    numbersep=5pt,                  
    showspaces=false,                
    showstringspaces=false,
    showtabs=false,                  
    tabsize=2
}
\begin{document}
\newcommand{\agoxrepo}{\url{https://gitlab.com/agox/agox}}
\newcommand{\agoxdocu}{\url{https://agox.gitlab.io/agox}}
\newcommand{\agoxdata}{\url{https://gitlab.com/agox/agox_data}}
\newcommand{\license}{GNU GPLv3}

\title[]{Atomistic structure search using local surrogate model}
\author{Nikolaj Rønne}
\author{Mads-Peter V. Christiansen}
\author{Andreas Møller Slavensky}
\author{Zeyuan Tang}
\author{Florian Brix}
\author{Mikkel Elkjær Pedersen}
\author{Malthe Kjær Bisbo}
\author{Bjørk Hammer}

\email{hammer@phys.au.dk}
\affiliation{Center for Interstellar Catalysis, Department of Physics and Astronomy, Aarhus University, DK-8000 Aarhus, Denmark}

\begin{abstract}
  We describe a local surrogate model for use in conjunction with global structure search methods. The model follows the Gaussian approximation potential (GAP)
  formalism and is based on a the smooth overlap of atomic positions descriptor with sparsification in terms of a reduced number of local
  environments using mini-batch $k$-means. The model is implemented in the Atomistic Global Optimization X framework and used as a partial replacement of the local
  relaxations in basin hopping structure search.
  The approach is shown to be robust for a wide range of atomistic system
  including molecules, nano-particles, surface supported clusters and surface thin films.
  The benefits in a structure search context of a local surrogate
  model are demonstrated. This includes the ability to transfer
  learning from smaller systems as well as the possibility to perform concurrent multi-stoichiometry searches.
\end{abstract}

\maketitle

\section{Introduction}
The use of electronic structure calculations has recently undergone dramatic changes driven by the introduction of machine
learning (ML) techniques used in the construction of potential energy surface (PES) surrogate models. 
Supervised machine learning regression methods  have
successfully been trained on large atomistic structure databases and used for the accurate and fast prediction of the PES.
Outstanding results have been obtained with kernel based methods\cite{bartok_gaussian_2010,chmiela_machine_2017, bartok_machine_2017,deringer_realistic_2018},
such as Gaussian Process Regression (GPR), as well as with deep neural
networks\cite{behler_generalized_2007,schutt_quantum-chemical_2017,schutt_schnet_2018,lubbers_hierarchical_2018,bogojeski_quantum_2020,zaverkin_fast_2021}.
Replacing the computational expensive electronic structure calculations, such as density functional
theory (DFT), with fast ML potentials has enabled hitherto inaccessible materials modeling.\cite{xie_ultra-fast_2021}
Example of this are longer molecular dynamics simulations providing
more reliable simulated vibrational spectra\cite{gastegger_machine_2017}, slower
cooling simulations providing more accurate insight in amorphous
solid phases\cite{jana_structural_2019} and more thorough global optimization searches
providing new structural models of e.g.\ point-defects \cite{arrigoni_evolutionary_2021}, surface reconstructions\cite{merte_structure_2022}
and supported clusters\cite{kolsbjerg_neural-network-enhanced_2018}.

The construction of ML potentials typically rely on the pre-construction of large and diverse
databases of atomic structures along with their target potential energies and forces. This is both time-consuming and
difficult since the database should be both diverse and an exhaustive sample of the chemical space of interest.
Several efficient schemes have been suggested for the construction of atomic structure databases by sampling the PES,
such as random structures searches\cite{deringer_data-driven_2018}, genetic algorithm\cite{bisbo_efficient_2020}, molecular dynamics sampling
\cite{behler_constructing_2015,li_molecular_2015,jinnouchi_--fly_2019}, meta-dynamics sampling\cite{xu_accelerating_2021}, density guided approaches\cite{schmitz_gaussian_2020}, Monte Carlo techniques\cite{loeffler_active_2020},
simulated annealing\cite{timmermann_data-efficient_2021} or local optimization techniques\cite{lin_automatically_2020}. Common for
all approaches is that they replace some of the computationally expensive target potential evaluations with
cheaper machine learning surrogate models.

The use of ML regression techniques for materials modelling requires a transformation from the Cartesian
coordinates of the individual atoms of an atomic structure to feature vectors suitable for regression.
Typically this includes translational and rotational invariance of the representation as well as permutational invariance of the atomic types.
The features can either be atomic descriptors representing
the local environment of each atom or a global descriptor representing the structural configuration of the entire
atomistic structure. Many approaches have been proposed for both local\cite{behler_atom-centered_2011,bartok_representing_2013,faber_alchemical_2018,huo_unified_2018}
and global\cite{valle_crystal_2010,rupp_fast_2012,montavon_learning_2012,faber_crystal_2015} descriptors.
A recent review of the topic can be found in Musil et al.\cite{musil_physics-inspired_2021}.

A multitude of global structure search methods have been proposed over the years. From the simple
and unbiased random structure search\cite{pickard_ab_2011}, where a random structure is generated and then relaxed
to a local minimum in the target potential, to the more elaborate genetic algorithms\cite{deaven_molecular_1995,ljohnston_evolving_2003,vilhelmsen_genetic_2014},
Other notable algorithms include basin hopping\cite{wales_global_1997}, simulated annealing\cite{kirkpatrick_optimization_1983},
minima-hopping\cite{goedecker_minima_2004} and particle swarm algorithms\cite{wang_crystal_2010,lv_particle-swarm_2012}.
Recently, much progress has been gained by leveraging machine learning (ML) techniques both for the screening
of candidates and data-driven generation of candidates\cite{jacobsen_--fly_2018,paleico_global_2020,simm_symmetry-aware_2020,kaappa_global_2021,gebauer_inverse_2022}
as well as local relaxations using ML potentials.\cite{schmitz_gaussian_2018,garijo_local_2019,yang_machine-learning_2021}

We have earlier introduced the Atomistic Global Optimization X (AGOX) code as a
customizable and efficient global structure optimization code.\cite{christiansen_atomistic_2022}
In this work, we implement a local GPR model based on the smooth overlap of atomic positions\cite{behler_atom-centered_2011}
(SOAP) representation as a surrogate model in the AGOX framework.
This is used to partly replace the structure relaxations in a parallel tempering basin hopping (PT-BH) structure search.
The model is trained on-the-fly during the search thereby enabling immediate feedback from the
previously evaluated structural candidate and efficiently sparsified to reduce the computational cost of
both training and prediction. The use of a local
descriptor based model opens for the possibility of 
transfer learning from pre-evaluated data. This is explored by training on data gathered
from smaller but similar systems and by performing concurrent searches for a multitude of
stoichimetries, where the model data is shared between the searches.
Compared to earlier work on local GPR models\cite{deringer_machine_2017,jinnouchi_--fly_2019,vandermause_--fly_2020}
our emphasis is on efficient training such that to enable effective on-the-fly learning
through the use of a mini-batch $k$-means sparsification, only training on energies as well as excluding
the training of two- and three-body models. We show that this simplistic approach is robust for a number
of systems ranging from nano-particles to surface thin films. The search for two-dimensional $\text{C}_5\text{NH}_5$ and $\text{C}_{30}$-cluster are
benchmarked against a PT-BH search using a global GPR model. An illustration of the robustness of the method is given by the search
for a number of diverse systems already tackled in the literature.
The systems are: $\text{Ti}_{13}$-cluster, bulk $\text{B}_{12}$, two-dimensional $\text{CoB}_{18}^-$,
$\text{Ag}_{12}\text{S}_6$-cluster, $(\text{MgSO}_3)_4$ and $\text{Cu}_{10}$
on a $\text{ZnO}(10\bar{1}0)$ surface.
Furthermore, we study the oxidation of a Ag(111) surface using a concurrent multi-stoichiometry BH search. Twelve different stoichiometries are
searched for simultaneously, where each search shares a common database and model thus exploiting online transfer learning between the different
sized systems. 

The paper is outlined as follows.
First, an introduction of the Gaussian approximation potential (GAP) formalism is provided along with our specific implementation choices.
Second, the ML-enhanced PT-BH search method is introduced in terms of the AGOX framework.
Third, a comparison between local and global surrogate model relaxations in conjunction with the PT-BT search method presented.
Fourth, a demonstration of the ML-enhanced PT-BH is performed for a number of
previously studied systems.
Last, the oxidation phase of a Ag(111) surface is studied using a concurrent multi-stoichiometry BH search.

\section{Method}
\subsection{Local Gaussian process regression model}

\begin{figure}
  \centering
  \includegraphics[width=0.45\textwidth]{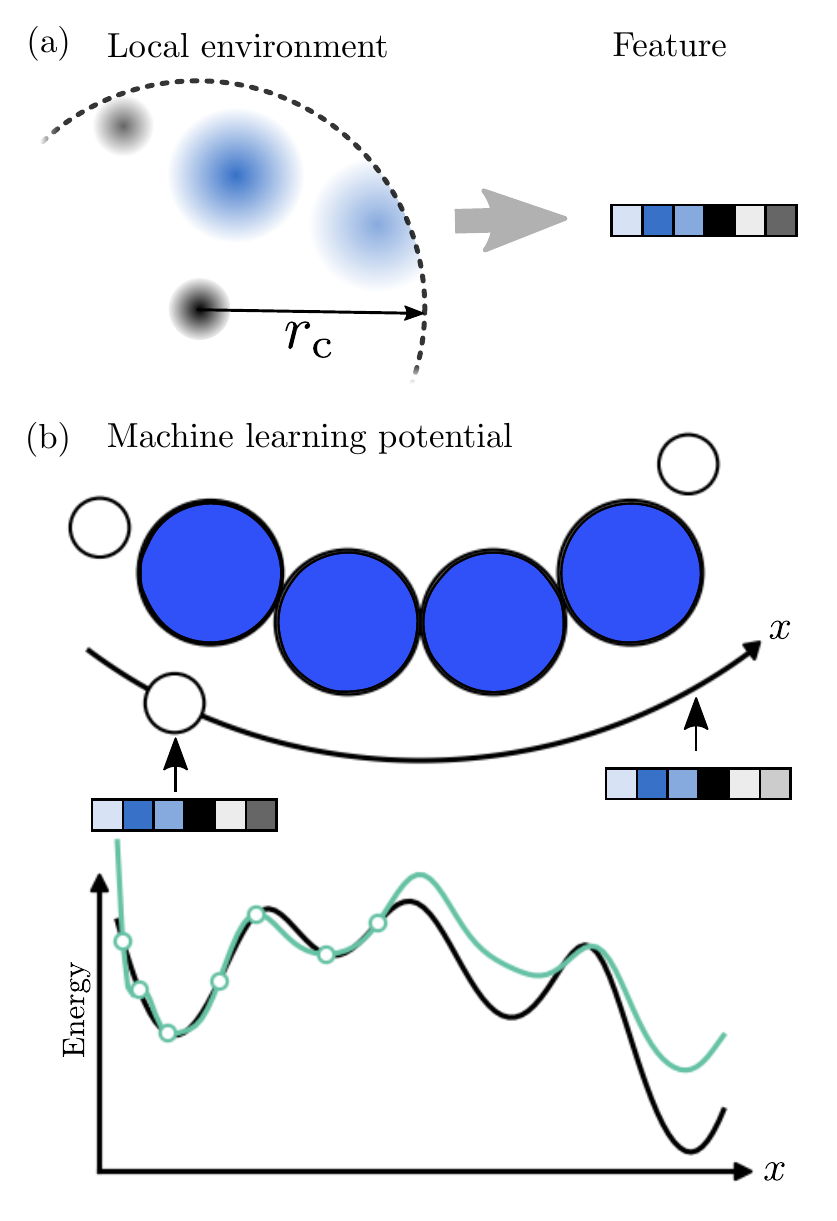}
  \caption{(a) The transformation from an atomic environment into a local feature representation.
    (b) Parameterised path of moving one hydrogen along the backbone of a molecule. Green outline dots
    represent structures included in the training of a local model. The green line represents prediction of the model.
    Albeit a schematic, the data in the figure are indeed constructed following the local GPR model outlined in this section.
  }
  \label{fig:intro}
\end{figure}

Kernel methods, such as Gaussian Process, rely on the transformation of the initial representation of the data
by a kernel function which enables fitting in an implicit multidimensional feature space. The kernel function can be
interpreted as a similarity measure between data points, and such models learns properties by measuring similarity to
known training examples through the kernel function. A Gaussian process is defined by its kernel and mean function which
together constitute a distribution of learnable functions prior to observations. By conditioning on observations a
predictive distribution arises with its mean taken as the model prediction. Typically a noise term is included
which expresses the amount of uncertainty, both from observation noise and any representation errors, of the observations.

Fig. \ref{fig:intro} schematically illustrates the concepts of representation and a local machine learned model for the generic compound
$\text{X}_3\text{Y}_3$. 
Fig. \ref{fig:intro}(a) shows the local environment of a white atom and the transformation to a local feature representation.
Fig. \ref{fig:intro}(b) shows the total energy prediction of a local model for the parameterised path of moving one white atom along the molecule backbone.
The green points indicate data used for the training of a local surrogate model, with the green line the model prediction.
Notice the promising extrapolation far beyond any training data on the right side of the parametrised path.

\begin{figure*}
  \centering
  \includegraphics[width=0.95\textwidth]{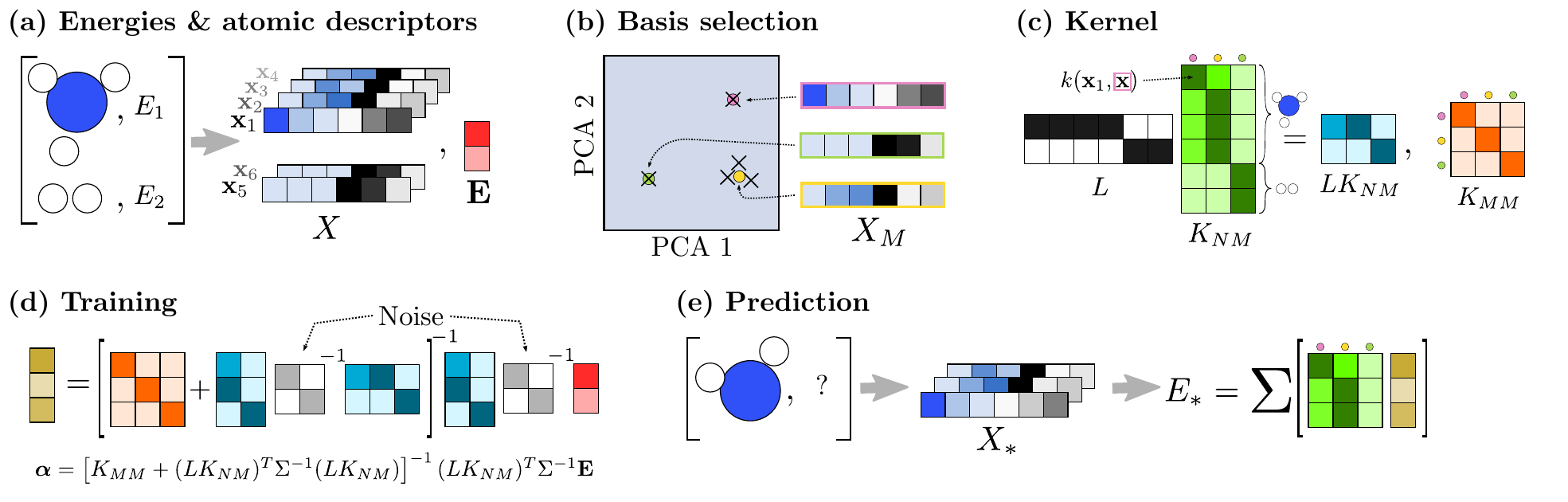} %model-figure-crop.pdf
  \caption{
    The training and prediction with a local GPR model. (a) For each structures in
    the training database the atomic descriptors are calculated. (b) Based on all the local atomic features, the
    sparsification is performed by selection a predefined number of basis feature vectors. In this example using $k$-means
    clustering of the features with $M=3$. (c) The kernel between the basis feature vectors and
    all the local features stemming from the structures are calculated and transformed with $L$ defined by
    eq. (\ref{eq:energy-transformation}). Furthermore, the kernel matrix between the basis vectors is calculated.
    (d) The $\boldsymbol{\alpha}$ fitting weights are found by solving the matrix equation.
    (e) The prediction of a new structure consists of calculating the
    atomic features. Calculating the kernel between each feature and the basis feature vectors and multiplying
    with the fitting weights. This gives the local energies, which are summed to give the total energy prediction.
  }
  \label{fig:model}
\end{figure*}

The GAP formalism extends the GPR formulation to encompass learning total energies through a number of local atomic representation without
explicitly defining a global descriptor. The fundamental assumption of the GAP formalism is that the potential energy of an
atomic structure is local, such that the total energy can be decomposed into the non-physical local energies obeying
\begin{equation}
  E = \sum_n \varepsilon_n.
\label{eq:total-energy}
\end{equation}
This is not generally true, since long-range electrostatic effects and
quantum mechanical phenomena (e.g.\ non-local extension of orbitals and fluctuating
dipole interactions) cannot be captured, but it has proven a very good approximation for many systems.
Thus, a GAP should
\textit{learn} to predict the local energies of a given structure based on the corresponding atomic descriptors, and the prediction
of the total energy of a given structure is simply the sum of local energies. Although this sounds simple, 
local energies are not present for ab-initio methods such as DFT, and thus it is not possible to use a standard GPR formulation, where
the number of dependent and independent variables must match.

The solution to this is through the decomposition of the total energy given by eq. \ref{eq:total-energy}, which for a number
of total energies corresponding to atomistic structures can be written as
\begin{equation}
  \mathbf{E} = L \boldsymbol{\varepsilon},
  \label{eq:energy-transformation}
\end{equation}
where $\boldsymbol{\varepsilon}$ are all the unknown local atomic energies. The matrix $L_{ij}$ is one if local energy $j$
corresponds to an atom in structure $i$ and otherwise it is zero. An example of the L matrix can be seen in fig. \ref{fig:model}(c).
In the GAP formalism the transformation matrix $L$ is
used to relate the local kernel matrices between the atomic descriptors to the corresponding total energies.

The number of atomic environments, $N$, grows very quickly with the number of training structures and thus the $\mathcal{O}(N^3)$ scaling for GPR training
becomes a computational bottleneck.
This is overcome by introducing a sparse GPR where
a user-defined number, $M$, of basis points in feature space, are used as representative points for all training data.
This reduces the time complexity for training to $\mathcal{O}(NM^2)$, as well as reducing prediction times complexity to being linear in $M$.
Hence, choosing a small number of representative basis features drastically reduces the computational demands of the model.
The sparsification is furthermore supported by the assumption that local environments are
repetitious in atomic structures, and thus redundant basis points can be removed without reducing the quality of the model.
Naturally over-sparsifying will lead to a worse fit, since areas of feature space present in the training data are
underrepresented in the sparse basis.

The kernel function quantifies the similarity between local atomic environments. The cornerstone of the
sparse GPR method is to only calculate the similarity between all local environments in the training data, $X_N$, and the basis points, $X_M$,
resulting in the sparse kernel matrix $K_{MN}=K_{NM}^T = K(X_M,X_N)$.
Using the energy decomposition given in eq. \ref{eq:energy-transformation} as well as the sparse kernel matrix $K_{\text{NM}}$,
the training of a local sparse
GPR reduces to solving eq. \ref{eq:GAP} for the training weights $\boldsymbol{\alpha}$.

\begin{align}
  \begin{split}
  \boldsymbol{\alpha} = \left[K_{MM} + (LK_{NM})^T \Sigma^{-1} (LK_{NM}) \right]^{-1}& \times\\
  (LK_{NM})^T &\Sigma^{-1} \mathbf{E},
  \label{eq:GAP}
  \end{split}
\end{align}
with $\Sigma$ being a diagonal noise matrix where $\Sigma_{ii}$ sets the noise value for local environment $({X_N})_i$.
In our implementation, we train the model using a $QR$-factorization of the kernel matrix part and
then solve the matrix equation using least squares. A further
regulariser in the form of a diagonal matrix with a small number on the diagonal is added before
factorization to reduce numerical instability. The steps
needed in order to train the model are illustrated in fig. \ref{fig:model}(a)-(d). In the limit of isolated basis point, the fitting weight $\boldsymbol{\alpha}_i$ can be interpreted as
the local energy contribution for \textit{looking like} basis point $i$ when measured in kernel similarity between atomic descriptor environments.

The prediction of the total energy is depicted in \ref{fig:model}(e) and is given by
\begin{equation}
  E = \sum_i  \mathbf{k}(X_M,\boldsymbol{x}_i)^T \boldsymbol{\alpha}, % check how i should with this, so it fits with everything else
 \label{eq:prediction} 
\end{equation}
where the sum is over all local environments in the structure, $K(X_M,\boldsymbol{x}_i)$ is the kernel similarity
between the feature vector for the local environment of atom $i$, $\boldsymbol{x}_i$, and
the matrix of all basis feature vectors, $X_M$.
Due to the sparsification this only includes $M$ kernel evaluations no matter the amount of the training data.
For an in-depth review of the GAP formalism as well as several use cases see Deringer et al.\cite{deringer_gaussian_2021}.

How to choose the basis points in the most suitable manner is still an open question. Previously this has been done using
$k$-means\cite{szlachta_accuracy_2014} and using the CUR algorithm\cite{deringer_machine_2017}. In this work, we have
chosen a mini-batch version of $k$-means\cite{sculley_web-scale_2010} that allows for the efficient re-training of the model during
a structure search.
The mini-batch $k$-means divides all training data into a number of batches. Each batch is then used to update the cluster centers
by gradient descent, such that all training data has appeared once in each training iteration.
In a typical structure search setting, where 
only a few structures are added to the training data per search iteration, only a single $k$-means training iteration is performed restarting
from the previous cluster centers.
In this work, we have chosen to use a Gaussian kernel together with the SOAP descriptors as implemented in the DScribe package\cite{himanen_dscribe_2020}.
The kernel is given by
\begin{equation}
  k(\mathbf{x}, \mathbf{x}') = A \exp \left( -\frac{d(\mathbf{x}, \mathbf{x}')^2}{2l^2}\right),
\end{equation}
where $d(\mathbf{x}, \mathbf{x}')$ is the Euclidean distance between atomic feature $\mathbf{x}$ and $\mathbf{x}'$.
Though our implementation allows the usage of all kernels buildable from the Scikit Learn package\cite{pedregosa_scikit-learn_2011}.
Additionally, an analytical repulsive term is added to the prediction of the energy to avoid atoms locally relaxing
to very short bond distances.\cite{deringer_machine_2017} The specific model and representation hyperparameters used for each system can be found in the supplementary material.

\subsection{Machine learning enhanced parallel tempering basin hopping}
\begin{figure}
  \centering
  \includegraphics[width=0.4\textwidth]{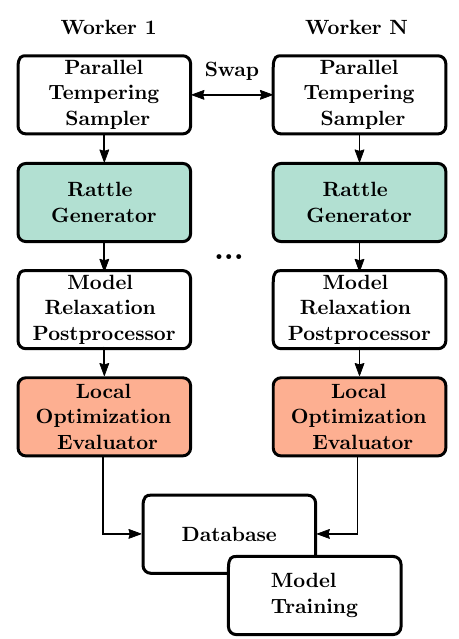}
  \caption{
    Flowchart of the ML-enhanced PT-BH search algorithm. Swaps between workers are carried out according to
    the acceptance criterion given by eq. \ref{eq:swap}.
  }
  \label{fig:GO}
\end{figure}

Now we introduce of the global structure optimization algorithm used in this work.
A ML-enhanced PT-BH structure search algorithm is proposed as an efficient global optimization (GO) algorithm
exploiting surrogate model relaxations in order to reduce the number target potential single point evaluations. 
In basin hopping a single structural candidate is evolved through a stochastic
rattle operation followed by a local relaxation. We partly replace the local relaxation by
a surrogate relaxation and only a few local relaxation steps are taken in the target potential. Hereby, the
computational cost of the search is significantly reduced. The rattled and relaxed candidate replaces its
parent structure as the starting point for the next iteration if the Metropolis criterion is fullfilled.
The acceptance criterion $A$ is given by
\begin{equation}
  \label{eq:metropolis}
  A = \text{min} \{ 1, \exp(\beta [E_{\text{parent}}-E_{\text{rattled}}]) \},
\end{equation}
where $\beta = 1/k_BT$ with $T$ a search hyperparameter, that determines the likelihood with which a less stable
structure can replace its parent. If the newly generated structure is not accepted the parent remains the starting
point for a new stochastic rattle operation followed by the relaxation. Furthermore, a parallel tempering scheme is used where several basin hopping
workers at different temperatures are run simultaneously. The workers are then allowed to swap their parent
structures if a Metropolis criterion similar to eq. \ref{eq:metropolis} is fullfilled. Specifically, the acceptance criterion for swapping
between two workers is given by
\begin{equation}
  A = \text{min} \{ 1, \exp([\beta_i - \beta_j] [E_{i}-E_{j}]) \},
  \label{eq:swap}
\end{equation}
where the indices $i$ and $j$ refers to two workers. By searching with a number of workers with varying temperatures
both high energy regions can effectively be explored as well as low energy regions can be exploited for an overall increased
efficiency.\cite{christiansen_atomistic_2022} A flowchart of the search method is shown in fig. \ref{fig:GO} following the AGOX modules naming.

All searches are performed in the AGOX framework with DFT using the
GPAW\cite{mortensen_real-space_2005,enkovaara_electronic_2010} code
except for the case of $\text{CoB}_{18}^-$-cluster search which is performed in DFT using the ORCA code.\cite{neese_orca_2012}
All DFT evaluations use the Perdew-Burke-Ernzerhof functional.\cite{perdew_generalized_1996}.
For the specific implementation details for each system see the supplementary material.

\section{Model benchmark}

\begin{figure*}
  \centering
  \includegraphics[width=0.95\textwidth]{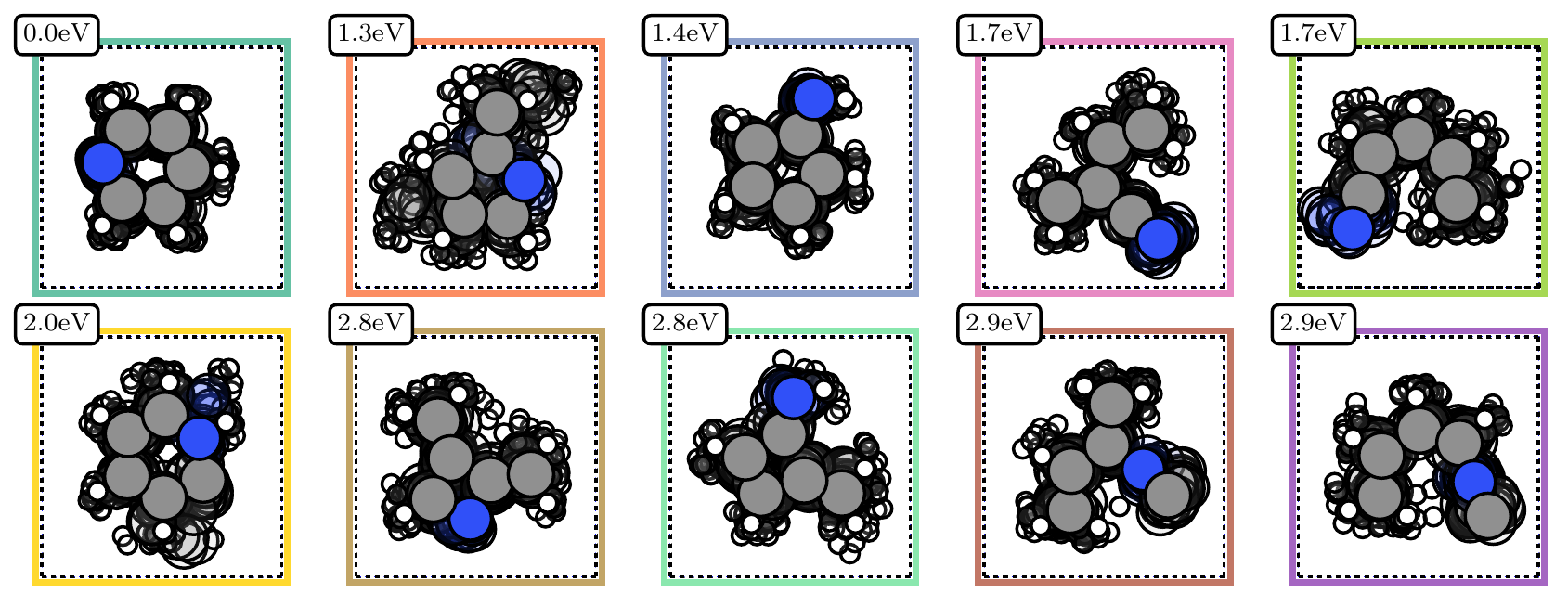}
  \caption{The ten two-dimensional conformers of $\text{C}_5\text{NH}_5$ and the MD structures used as test data.
    The transparent atoms behind each conformer indicate the variety of structures in the test data stemming from the MD runs.
    The border coloring is used to identify which conformer the training and test data refers to in the fig. \ref{fig:parity}. 
  }
  \label{fig:md-data}
\end{figure*}

We start by demonstrating the performance of the local GPR model and its ability to transfer knowledge across
stoichiometries on the standard energy regression problem. This is only an indication of the models
capabilities when used in an active learning setting, such as a structure search where the model is trained
on-the-fly. The system chosen is
two-dimensional $\text{C}_5\text{NH}_5$, since it constitutes a multi-species system with high enough complexity to be challenging, while
also being easy to depict visually. Furthermore, $\text{C}_4\text{NH}_5$ is used to test the transferability of
the model.
Constant energy molecular dynamics (MD) simulations at $4000$K starting from ten distinct low energy conformers
of $\text{C}_5\text{NH}_5$ are sampled to create
a training and test dataset. The ten conformers and the data present in the test dataset are visualised
in fig. \ref{fig:md-data}. 80 structures for each conformer are selected randomly for the training dataset and
20 structures from each for the test dataset.

\begin{figure*}
  \centering
  \includegraphics[width=1.0\textwidth]{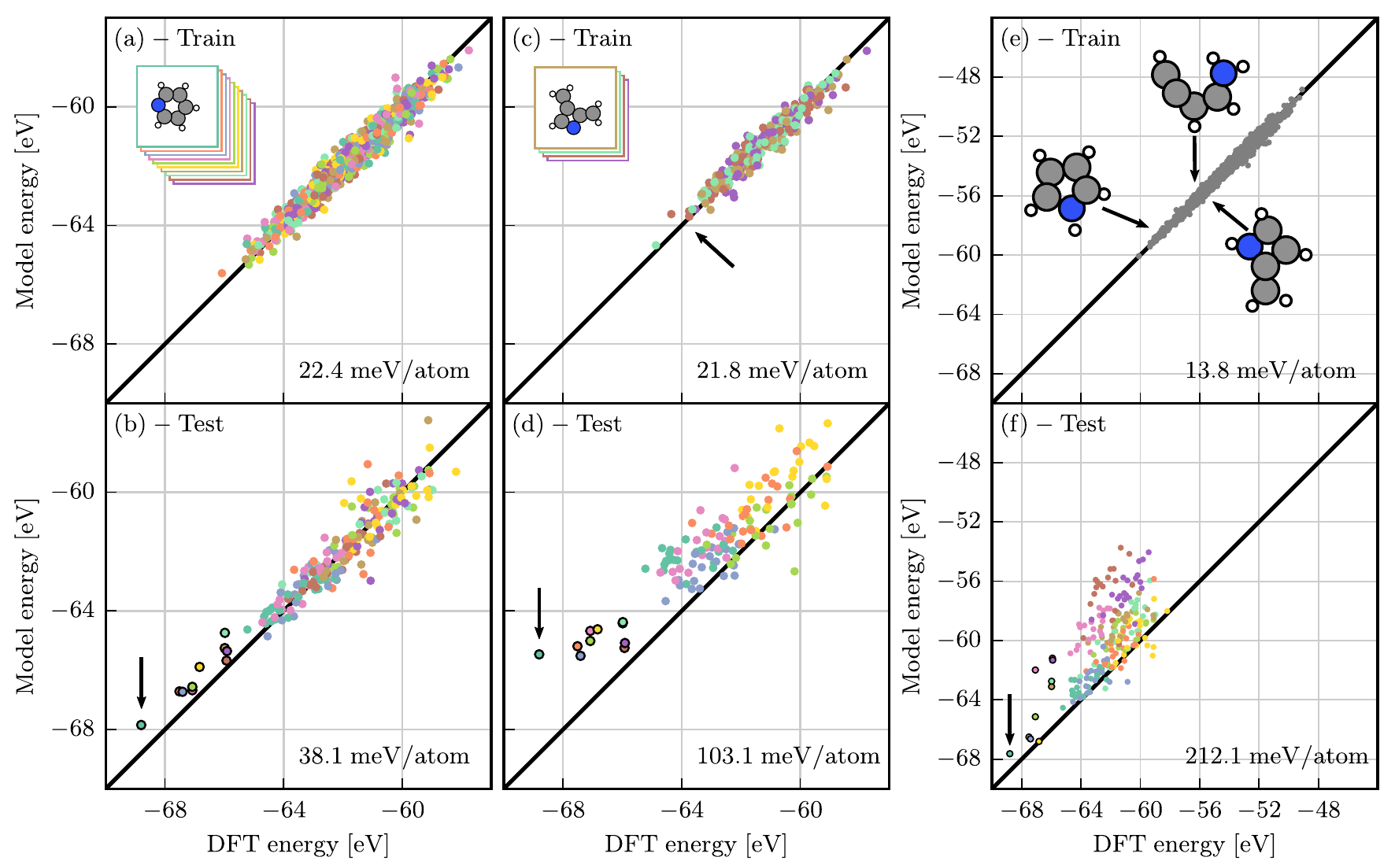}
  \caption{(a,b) The train-test energy predictions for the model trained on all the training data. Points with
    a black outline are the predictions on the ten unmodified conformers. The most stable conformer is
    also predicted as the most stable test structure as indicated by the arrow.
    (c,d) The train-test predictions for only training on the part of the training data belonging to the MD data from the
    four highest energy conformers.
    (e,f) The train-test predictions for training on $\text{C}_4\text{NH}_5$ data and testing on the $\text{C}_5\text{NH}_5$
    test data. Most structures in the test data are predicted far lower in energy than anything in the training dataset owing
    to excellent transferability.
    For each prediction the mean absolute error between the prediction and the DFT energy is stated. 
  }
  \label{fig:parity}
\end{figure*}

Fig. \ref{fig:parity}(a,b) shows the train-test error, where the model is trained on the entire MD training dataset.
The hyperparameters for the kernel length-scale, $l$, and the model noise are optimised with a grid search by $5$-fold cross validation.
The parameters with lowest mean absolute error upon validation are used, and
the model is trained on the entire training dataset.
The parameters of the SOAP descriptor are kept fixed throughout with $n_{\text{max}}=3$, $l_{\text{max}}=2$, $\sigma=1$
and a cutoff radius of $3$Å as well as a polynomial weight function as implemented in the DScribe package.
Fig. \ref{fig:parity}(b) demonstrates that the model is able to
interpolate the diverse set of structures present in the test data.
It is also worth noting that the model is able to predict energies
much lower than anything it has seen in the training data, and that these predictions are very reasonable, with
the lowest energy structure being correctly predicted to be the pyridine structure as indicated by the arrow.

Fig. \ref{fig:parity}(c,d) shows the train and test data energy predictions for a model only trained on the part of the
training data which belongs to the four highest energy conformers. The missing low energy conformers are pointed out in
fig. \ref{fig:parity}(c) with an arrow. 
Hence, no five or six ringed structures are included in the
training data. The test predictions are on the same data as in fig. \ref{fig:parity}(b) excluding structures
from the highest four energy conformers. The model fit is worse than in the previous case, which is expected from
the higher degree of extrapolation required. Despite this, the model is able to give good prediction for the test
structures, and the lowest energy conformers are reasonably well predicted.

One of the important features of a local model is its ability to transfer knowledge across stoichiometries by only
relying on local feature information. This enables pre-training models on more accessible data typically from smaller systems
and then using the model on larger or more complex systems. As a simple demonstration of the transferability of the model, we train
a model on a diverse dataset of $\text{C}_4\text{NH}_5$ structures. This model is then used to predict on the
$\text{C}_5\text{NH}_5$ test set. The predictions are shown in fig. \ref{fig:parity}(e,f) along with the
prediction errors. The model predictions are reasonable with especially branched structures being ill-predicted,
whereas ring structures are well predicted. This is a natural consequence of no structures in the training data showing
motifs with a branching carbon backbone, thus nothing explicit has been learned about such environments.
Again, it can be noted that the model is able to predict the low energy
conformers with high accuracy, and that their individual ordering of energies are almost correct.

This benchmark on a simple but diverse dataset stand as a testament to both the extrapolative power
of the local GPR model as well as its ability to transfer knowledge between stoichiometries.
In a GO search context, this allows us to solve smaller problems and transfer the data to help
improve the performance of a search on larger problems. 

\section{$\text{C}_5\text{NH}_5$ and $\text{C}_{30}$ search}
We now move on to the application of the local GPR model in a GO setting.
A two-dimensional search for $\text{C}_5\text{NH}_5$ where the global minimum (GM) is the heterocyclic molecule pyridine is
carried out with both a global and a local GPR model as the surrogate model used in the ML-enhanced PT-BH search strategy.
Two choices of global descriptors are compared by using either an averaged SOAP descriptor or the Valle and Oganov fingerprint descriptor\cite{valle_crystal_2010}
for the global GPR model searches.
For the local GPR model only the SOAP descriptor is used, but transfer of either the $\text{C}_4\text{NH}_5$ GM
or the entire $\text{C}_4\text{NH}_5$ training dataset used previously is presented.
Comparison is made by performing 50 independent searches and recording at which iteration each search
is successful, which in this case is defined by finding the GM pyridine structure.
The accumulative sum of the success-count as a function of number of single point DFT evaluations then results in a
statistical measure of the performance of a specific search method. See Christiansen et al. for a more detailed explanation of success-curves.\cite{christiansen_atomistic_2022}

\begin{figure}
  \centering
  \includegraphics[width=0.5\textwidth]{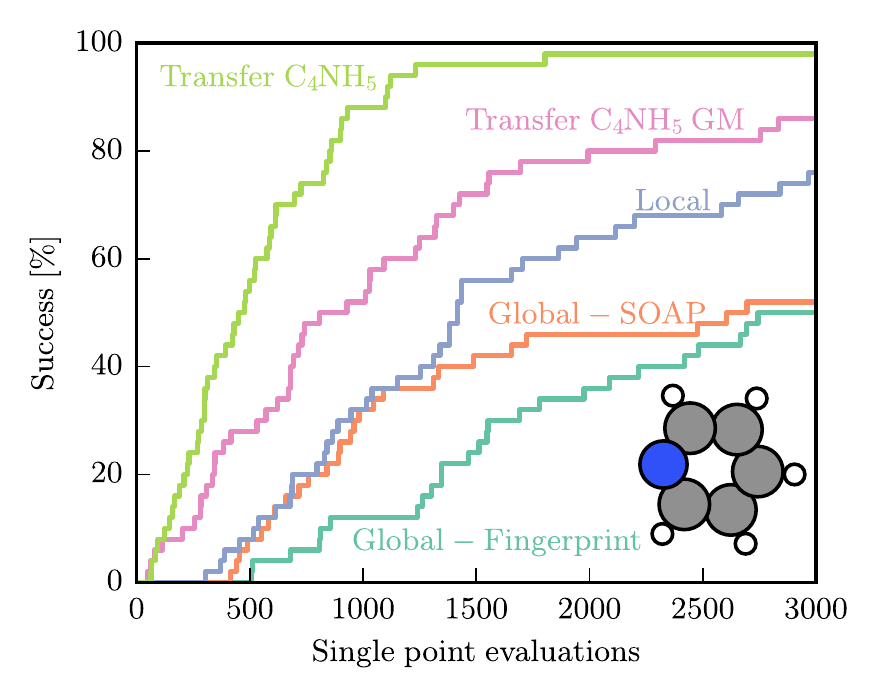}
  \caption{
    Success as a function of number of search single point evaluations for the two-dimensional $\text{C}_5\text{NH}_5$ structure search.
    The GM pyridine structure is shown in the bottom corner. 
  }
  \label{fig:C5NH5-success}
\end{figure}

Fig. \ref{fig:C5NH5-success} shows the success curves for the five different search settings. As is evident, an improvement is
gained by using a local GPR model compared to a global GPR model regardless of which representation is used with the
global model. Another improvement in the efficiency of finding the GM is gained by biasing the local model by including
the $\text{C}_4\text{NH}_5$ GM pyrrole structure in the training data from the start of the search. Finally, transferring
more $\text{C}_4\text{NH}_5$ data than purely the GM increases the performance further. This can be ascribed to the
much more detailed model PES trained from the start of the search already before any data is gathered during the search.
Note that the SOAP descriptors used for the global GPR model are different from the local GPR model. Several sets of parameters for the global SOAP descriptor
have been tested and the best choice is shown here.

\begin{figure}
  \centering
  \includegraphics[width=0.5\textwidth]{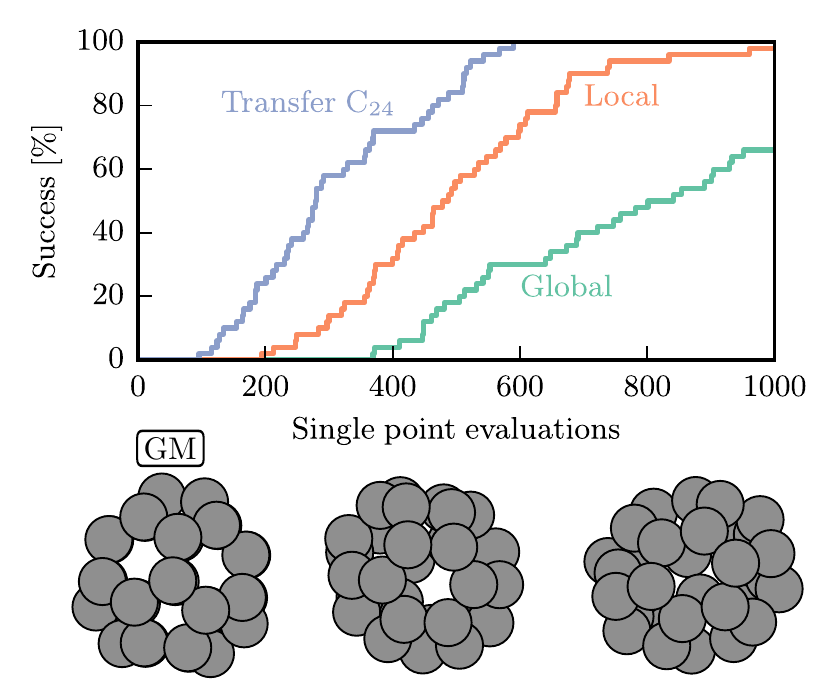}
  \caption{
    Structure search for $\text{C}_{30}$-clusters using a global GPR model as compared to
    a local GPR model with and without transfer data. The three distinct clusters within $0.1$eV
    of the GM are shown.
  }
  \label{fig:C30-success}
  
\end{figure}

We present another comparative study between the global and local GPR models for
a three-dimensional $\text{C}_{30}$-clusters search. This is shown in fig. \ref{fig:C30-success}.
Again, the local GPR model clearly outperforms the global GPR model using the
fingerprint descriptor. A further increase in success is gained by including the four lowest energy structures from $\text{C}_{24}$-clusters.\cite{bisbo_global_2022}

The capabilities of any machine learning potential is limited by the representation of the atomistic
structures. The SOAP descriptor allows for the adjustment of the representation to suit the specific use case through
its parameters such as the width of the Gaussian broadening of the atomic neighbour densities and the cutoff. 
The optimal representation-parameters for a given system is often difficult to guess and typically depends on the use case.
In a structure search setting where the data for the model is actively being gathered during the search, both very diverse and sometimes
irrelevant data are collected during the search.
Furthermore, an effective model in an active learning setting, such as the on-the-fly learning of a model in a structure search setting,
is balanced with respect to being confident about stable configurations outside its training data.
If the models keeps predicting false minima in the surrogate potential a search will continue producing such unreasonable structures.
On the other hand, if the model is not able to extrapolate away from its training data no exploration of the PES will occur
during the search. In both cases the search will stagnate and not be very effective.
The representation can affect this compromise between the extrapolation and interpolation qualities of the model.

\begin{figure}
  \centering
  \includegraphics[width=0.5\textwidth]{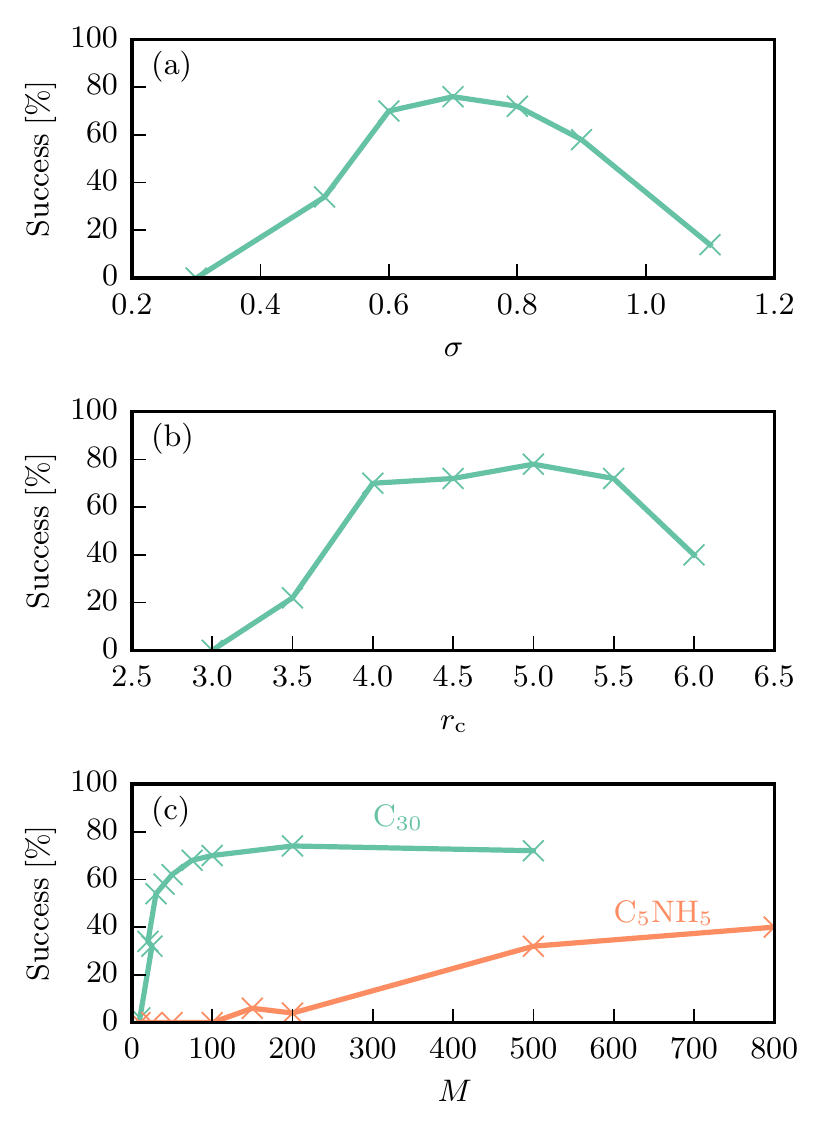}
  \caption{
    Search success after 600 single point evaluations with varying representation parameters.
    In (a) the $\sigma$ parameter of the SOAP descriptor is varied. In (b) the cutoff of the SOAP descriptor is varied.}
  \label{fig:parameter-search}
\end{figure}

Fig. \ref{fig:parameter-search} shows the success after 600 single points evaluations for a $\text{C}_{30}$-cluster search
with varying SOAP parameters. Specifically, the Gaussian broadening of the atomic neighbour densities, $\sigma$, in (a) and the cutoff $r_c$ in (b).
Small $\sigma$ values makes the representation very distinctive, which hinders the extrapolative
capabilities of the model and thereby makes the search stagnate in high energy configurations and hence poor success is reached.
On the other hand, too large values for $\sigma$ also leads to poor success. This can be attributed to the reduced distinctiveness
of the representation, which leads to
difficulties in resolving the optimal structural configurations. A similar trend is seen for the effect of the cutoff of the
SOAP descriptor. Here, a too short cutoff will lead to the reduced distinctiveness of the representation and thus the diminished resolution of
the model. A too long cutoff has the effect of increasing the number of distinctive local environments and thus making extrapolation difficult
as all environments appear distinct. As is evident from fig. \ref{fig:parameter-search}, the choice of parameters for the representation
is crucial for a search to be successful and a poor choice of parameters could lead to wrong conclusion when performing structure searches
for which the GM is unknown. Over-sparsifying the model by selecting to few basis points can also lead to an ineffective model in a GO setting.
Fig. \ref{fig:parameter-search}(c) shows the success after 600 single point evaluations for both $\text{C}_{30}$ and $\text{C}_5\text{NH}_5$.
It is evident that a lot more basis points is needed to resolve the multi-species $\text{C}_5\text{NH}_5$ system than single-species
$\text{C}_{30}$ system.

\section{Structure search examples}
\begin{figure*}
  \centering
  \includegraphics[width=\textwidth]{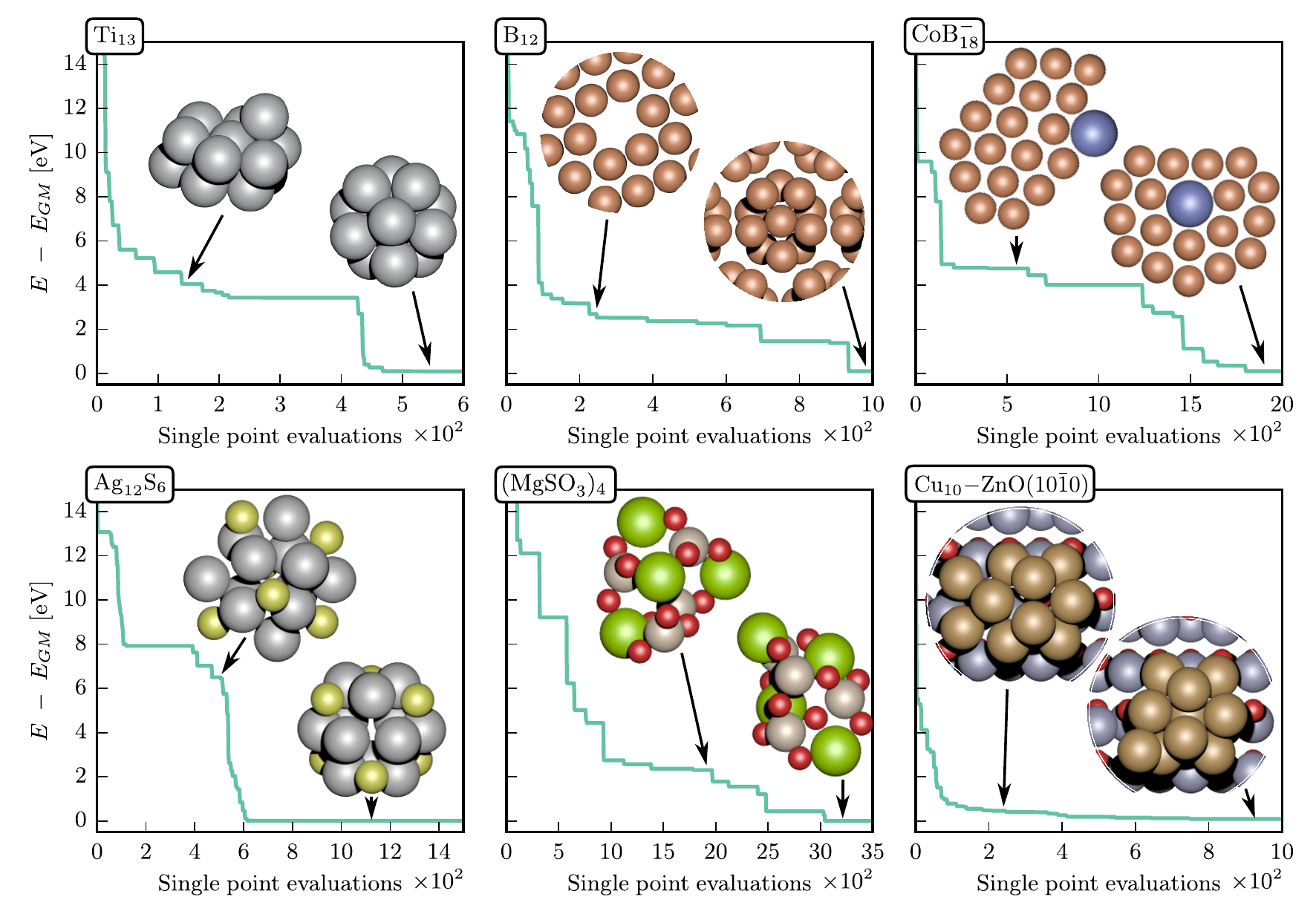}
  \caption{
    Examples of the ML enhanced PT-BH searches for
    a variation of different systems. For each system the energy progression relative
    to the GM energy is plotted. The GM structure along with an earlier
    candidate is shown.
  }
  \label{fig:examples}
\end{figure*}

After having seen how efficiently the ML-enhanced PT-BH search performs on the test systems we undertake searches
for systems already investigated in literature. A short overview of the chosen systems and with what methods they have been studied previously
is provided below.

The global optimization of $\text{Ti}_N$-clusters for $N=2-32$ has been studied by Lazauskas et al.\cite{lazauskas_thermodynamically_2018}
In their work, they use a genetic algorithm on a tight-binding interatomic potential with post-search analysis of low energy clusters using DFT.
We perform a search for the $\text{Ti}_{13}$-clusters using our ML-enhanced PT-BH algorithm.

Bernstein et al. have proposed the use of ML-enhanced RSS using a GAP potential for the search for
the bulk $\alpha$-rhombohedral boron structure.\cite{bernstein_novo_2019} This method utilizes the GAP model as
a surrogate landscape when performing RSS and iteratively improving the GAP model by selecting
geometrically diverse new training data using the CUR algorithm. Similarly, we perform
a ML-enhanced PT-BH search for the bulk $\alpha$-rhombohedral boron structure.

The GM structure of the planar Co-doped boron cluster, $\text{CoB}_{18}^-$, has been found by Li et al.
using their BH search method TGmin.\cite{li_planar_2016,chen_tgmin_2019} The search for the $\text{CoB}_{18}^-$
cluster is carried out using our ML-enhanced PT-BH method.
Unlike all other searches presented, this has been done using the ORCA\cite{neese_orca_2012} DFT package. 

Song and Tian have investigated $(\text{Ag}_2\text{S})_N$-clusters for $N=1-8$ using a generic algorithm and subsequently characterized 
the properties of the their GM.\cite{song_systematic_2019} We perform a structure search for the $(\text{Ag}_2\text{S})_6$ structure using our ML-enhanced
PT-BH method corroborating the GM found by Song and Tian.

Structure searches for nanosilicate pyroxene $(\text{MgSO}_3)_N$ for $N=1-10$ using a Monte Carlo basin-hopping method in a
specifically tailored interatomic potential followed by DFT characterization of the found GM has been carried out by Escattlar et al.\cite{escatllar_structure_2019}
We perform a search for the $N=4$ pyroxene cluster confirming the GM found by Escatllar et al.\cite{escatllar_structure_2019}

The investigation of $\text{Cu}_N$-clusters for $N=4-10$ on a $\text{ZnO}(10\bar{1}0)$ surface has previously been carried out
by Paleico and Behler.\cite{paleico_global_2020} In their work, they use a ML-enhanced genetic algorithm employing a high-dimensional
neural network potential. We have performed a search for the $N=10$ cluster on a frozen $\text{ZnO}(10\bar{1}0)$ surface
using our ML-enhanced PT-BH method confirming the GM found by Paleico and Behler.

Fig. \ref{fig:examples} shows the energy evolution of a single ML-enhanced PT-BH search using the local
GPR surrogate model for the six different systems studied in this section. In all cases the search algorithm successfully identifies the global
minimum structure as well as providing reasonable structural candidates during the search. This shows the applicability
of the method to a large variation of systems previously studied in the literature. For the specific search details see
the supplementary material.

\section{Silver-oxide concurrent search}
\begin{figure}
  \centering
  \includegraphics[width=0.45\textwidth]{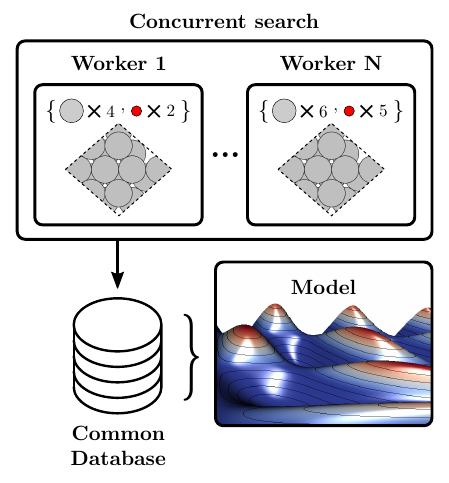}
  \caption{
    Illustration of the concurrent multi-stoichiometry search, where a number of workers search for structures with a
    varying number of silver and oxygen atoms, but share a common database and model.
  }
  \label{fig:concurrent-search}
\end{figure}

The oxidation of a $\mathrm{Ag}(111)$ surface has previously been studied both experimentally and theoretically \cite{schnadt_experimental_2009,mortensen_atomistic_2020}.
Thorough searches with varying stoichiometries along with thermodynamical calculations have successfully identified the
correct experimental structure of the $c(4 \times 8)$ phase, but relying on large amounts of computational resources, since each stoichiometry
is searched for completely independently from each other. We propose a concurrent searching scheme outlined in fig. \ref{fig:concurrent-search},
where a number $N$ concurrent searches are started simultaneously with a shared database and a shared model. Since the local GPR model does
not require the training structures to be of the same size, it is possible to train a common model based on all the data gathered in the $N$ searches, thus
enabling online transfer learning between multiple searches.
This has the benefit, that the model improves much faster as compared to a search with only one stoichimetry. Furthermore, motifs which
are difficult to construct in one stoichiometry might be easier in another and thus the model will quickly cover a large part local feature space.
The searches do not share structural candidates, but this could easily be implemented such that interesting structures with one stoichimetry can
be used in the search for a slightly different stoichiometry with the correct number of atoms added or subtracted.

\begin{figure*}
  \centering
  \includegraphics[width=1.0\textwidth]{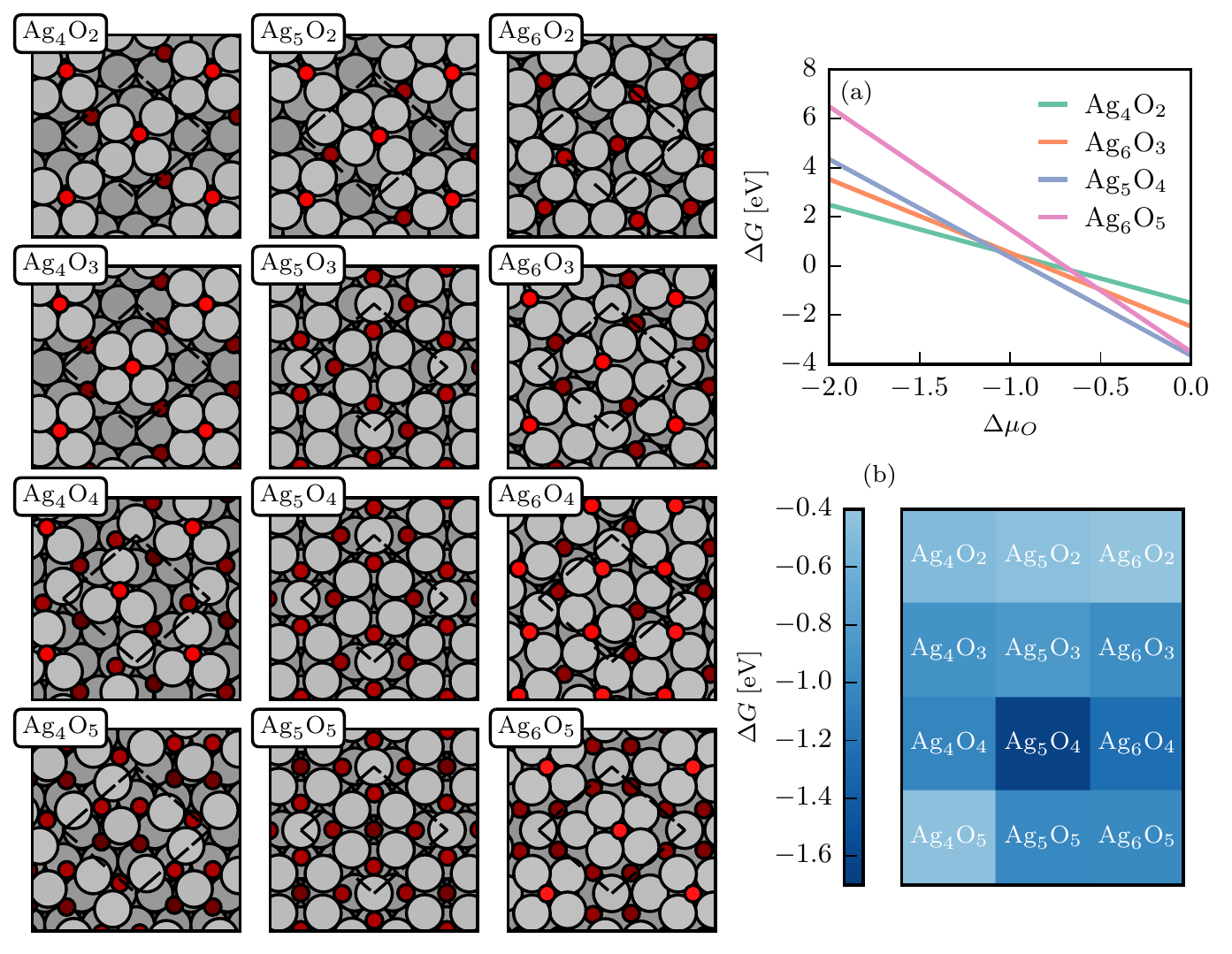}
  \caption{
    The most stable surface oxides for the twelve investigated stoichiometries along with the free energy $\Delta G$ as a function of
    $\Delta \mu_O$ for the three most stable silver oxides with varying oxygen coverage shown in (a) and a raster plot of the free energy
    $\Delta G$ for $\Delta \mu_O = -0.5\mathrm{eV}$ corresponding roughly to ambient conditions in (b).
  }
  \label{fig:AgxOy-search}
\end{figure*}

For the $\mathrm{Ag}(111)-c(4 \times 8)$ surface oxide search, twelve concurrent searches are started for $\mathrm{Ag}_X\mathrm{O}_Y$
with  $4 \leq X \leq 6$ and $2 \leq Y \leq 5$ using a two-layer Ag slab and thus covering the area
around the previously established most stable oxide phase, $\mathrm{Ag}_5\mathrm{O}_4$. The most stable structure for each stoichiometry
is plotted in fig. \ref{fig:AgxOy-search} corroborating the previous results using independent global structure searches. These are obtained
by relaxing the five most promising structural candidates on a five layer slab and plotting the lowest energy structure for each stoichiometry. 
It is worth noting, that several structures share the same motifs, such as the $\mathrm{Ag}_5\mathrm{O}_Y$ structures for $Y=3-5$, where the difference
is only the addition or subtraction of a single oxygen compared to $\mathrm{Ag}_5\mathrm{O}_4$. Triangular motifs with a single central on top oxygen is also apparent in several of
the most stable structures such as $\mathrm{Ag}_6\mathrm{O}_3$, $\mathrm{Ag}_4\mathrm{O}_4$ and $\mathrm{Ag}_6\mathrm{O}_4$. Furthermore, square motifs with a single central on top oxygen
are evident for lower coverage such as $\mathrm{Ag}_4\mathrm{O}_2$, $\mathrm{Ag}_5\mathrm{O}_2$ and $\mathrm{Ag}_4\mathrm{O}_3$.
Due to the repetition of stable motifs between stoichiometries the online transfer of knowledge between systems is
expected to have improved the search significantly.

Post search the most stable structures for each stoichiometry are compared by their Gibbs free energy according to\cite{reuter_composition_2003}
\begin{equation}
  \Delta G(T, p) = E^{\mathrm{DFT}} - E^{\mathrm{DFT}}_{\mathrm{slab}} - Y \mu_{\mathrm{Ag}} - X \mu_{\mathrm{O}}(T,p),
\end{equation}
where $E^{\mathrm{DFT}}$ is the DFT energy of the entire structure and $E^{\mathrm{DFT}}_{\mathrm{slab}}$ is the energy of the $\mathrm{Ag}(111)-c(4 \times 8)$
five layer slab. The chemical potential of Ag is calculated as the difference per atom between a six- and five-layer slab. The oxygen chemical potential
is calculated as\cite{reuter_composition_2003}
\begin{equation}
  \mu_{\mathrm{O}}(T, p) = \Delta \mu_{\mathrm{O}}(T, p) + \frac{1}{2}E_{\mathrm{O}_2}^{\mathrm{DFT}},
\end{equation}
where $\Delta \mu_{\mathrm{O}}(T, p)$ is the temperature and pressure dependent part of the chemical potential. The free energy for $\mathrm{Ag}_4\mathrm{O}_2$, $\mathrm{Ag}_6\mathrm{O}_3$,
$\mathrm{Ag}_5\mathrm{O}_4$ and $\mathrm{Ag}_6\mathrm{O}_5$ are shown in fig. \ref{fig:AgxOy-search}(a) for varying $\Delta \mu_{\mathrm{O}}(T, p)$. It is observed,
that for a wide range of $\Delta \mu_{\mathrm{O}}$-values the $\mathrm{Ag}_5\mathrm{O}_4$ stoichiometry is the most stable phase. Fig. \ref{fig:AgxOy-search}(b)
shows the free energy for the twelve investigated stoichiometries with $\Delta \mu_{\mathrm{O}} = -0.5\mathrm{eV}$ corresponding roughly to ambient conditions.\cite{mortensen_atomistic_2020}
The concurrent search method enabled by the use of a local model allows for effortless searches for the most stable phase when the experimental composition is
unknown.

\section{Conclusion}
We have introduced a local GPR surrogate model based on the GAP formalism and implemented it in the AGOX framework
for the use in conjunction with structure search methods. The local GPR model has been used as a partial replacement
for the local relaxations in BH structure search and it is shown that it is efficient and robust on a number of atomistic
systems. Furthermore, transfer learning has been successfully exploited both for pre-training on smaller system
and online transfer in concurrent multi-stoichiometry structure searches. 

\section{Acknowledgements}
This work has been supported by VILLUM FONDEN through Investigator grant, project no. 16562,
and by the Danish National Research Foundation through the Center of Excellence “InterCat” (Grant agreement no: DNRF150).
 
\section{Data availability}
The AGOX package is publically available at \agoxrepo \ under a \license \ license.
Documentation available at \agoxdocu. Data supporting the findings presented 
in this paper available at \agoxdata.

\section{References}
\bibliography{bib}

\end{document}